\documentclass[
superscriptaddress,
article
 amsmath,amssymb,
 aps,
12pt]{revtex4}
\usepackage{float}
\usepackage{natbib}
\usepackage{mathtools}
\usepackage{eucal}
\usepackage[labelformat=simple]{subcaption}
\usepackage[dvipsnames]{xcolor}
\usepackage[utf8]{inputenc}
\usepackage{amsmath}
\usepackage[T1]{fontenc}
\usepackage{amsmath,amsfonts,amssymb,amsthm,epsfig,epstopdf,url,array}
\usepackage{soul}
\usepackage{braket}
\usepackage{dcolumn}% Align table columns on decimal point
\usepackage{bm}% bold math
\usepackage{graphicx}
\usepackage{comment}
\usepackage[caption=false]{subfig}
\usepackage{lipsum}
\usepackage{soul}
\newcommand{\bea}{\begin{eqnarray}}

\newcommand{\eea}{\end{eqnarray}}

\usepackage{color}

\begin{document}

\title{Spectra and transport of probability and energy densities in a $\mathcal{PT}$-symmetric square well with a delta-function potential}
% Force line breaks with \\
\author{Francisco Ricardo Torres Arvizu}
\affiliation{Instituto de Ciencias Físicas, UNAM, Av. Universidad s/n, CP 62210 Cuernavaca Morelos, México}
\author{Adrian Ortega}
\affiliation{Departamento de Física, Universidad de Guadalajara, Blvd. Gral. Marcelino García Barragán 1421, C.P. 44430, Guadalajara, Jalisco, M\'exico}
\author{Hernán Larralde \footnote{Email: hernan@icf.unam.mx}}
\affiliation{Instituto de Ciencias Físicas, UNAM, Av. Universidad s/n, CP 62210 Cuernavaca Morelos, M\'exico}

%\date{\today}

\begin{abstract}

We study the spectrum, eigenstates and transport properties of a simple $\mathcal{P}\mathcal{T}$-symmetric model consisting in a finite, complex, square well potential with a delta potential at the origin. We show that as the strength of the delta potential increases, the system exhibits exceptional points accompanied by an accumulation of density associated with the break in the $\mathcal{P}\mathcal{T}$-symmetry. We also obtain the density and energy density fluxes and analyze their transport properties. We find that in the $\mathcal{P}\mathcal{T}-$ symmetric phase transport is efficient, in the sense that all the density that flows into the system at the source, flows out at the sink, which is sufficient to derive a generalized unitary relation for the transmission and reflection coefficients.

\end{abstract}

%\date{May 2022}

\maketitle

\section{Introduction}

Since its introduction by Bender and Boettcher \cite{bender1}, $\mathcal{P}\mathcal{T}$-symmetric quantum mechanics has attracted much interest due to the phenomena and applications that it can describe, see, for example \cite{Bagarello2016-BAGNHI, El-Ganainy, christodoulides2018parity, bender2019pt} and references therein.  One of the most studied aspects of non-Hermitian $\mathcal{PT}$-symmetric Hamiltonians is the appearance of exceptional points as a parameter of the system is varied. At such points, the $\mathcal{PT}$ symmetry of the solutions breaks, the eigenvalues become complex \cite{zyablovsky2014pt}, and two or more eigenvectors coalesce. As a result, the spectral decomposition of operators fails, and the Hamiltonian becomes defective \cite{kato2013perturbation, Heiss_2012, bender2019pt}. \\

To gain insight on the phenomenology that can arise in this kind of systems, it is desirable to have analytical models in which a system with a $\mathcal{P}\mathcal{T}$-symmetric potential can be solved analytically (see for example~\cite{Moibook2011} and references therein, for a general overview on non-Hermitian systems). For instance ~\cite{LevaiJPConfS2008} addresses central potentials while references below address one-dimensional square well and step potentials. 
However, only a few $\mathcal{P}\mathcal{T}$-symmetric potentials are solvable \cite{bender2019pt}, perhaps the most explored have been one-dimensional square wells in their distinct versions. 
The first system of this type was a complex extension of the finite potential well where its bound states were calculated \cite{pozo1}. Due to the simplicity of this kind of systems, they have been studied in frameworks such as pseudo-Hermitian quantum mechanics  \cite{Mostafazadeh1} as well as from the perspective of supersymmetry ~\cite{supersymmetric}, while other studies have focused on aspects such as the breaking of the $\mathcal{P}\mathcal{T}$-symmetry in a step potential~\cite{PTwellBreak}. 
Later, modifications with periodic boundary conditions \cite{Circle_one_step}, discrete systems \cite{Discrete_well, Barashenkov},  two coupled $\mathcal{P}\mathcal{T}$-symmetric square wells ~\cite{coupled-symmetricsquarewells}  and spatially antisymmetric~\cite{Quesne}  versions also have been investigated. Surprisingly, $\mathcal{P}\mathcal{T}-$symmetric finite wells in an infinite interval, whose Hermitian form is traditionally addressed in almost any standard mechanics course \cite{hermitianwell}, have barely been studied \cite{levai2018finite}. This model is interesting because it presents scattering states as well as an infinite set of bound state solutions \cite{levai2018finite}. 
On the other hand, other solvable models related to  $\mathcal{PT}-$symmetric square wells are systems containing delta potentials \cite{Znojil_2005, uncuPLA2006, barashenkovChapterinBook2016, Kovács_Lévai_2017, mostafazadeh2018scattering}. The mechanisms of the level crossings and of the $\mathcal{PT}$-symmetry breaking in these systems, as well as their scattering properties, have been widely studied \cite{Znojil_2005, zhang2017transport}, and applied in the context of Bose-Einstein condensation \cite{Dast_2013, Cartarius, cartarius2}. \\

Traditionally, the physical meaning of the imaginary part of the potential is that it acts as source or drain of probability; thus, even in stationary states, fluxes are present in the system.  Some studies on this subject have found that the transport of probability density in the $\mathcal{P}\mathcal{T}$-symmetric phase in tight binding models is efficient, this is to say that in the $\mathcal{P}\mathcal{T}$-symmetric phase the gain and loss of probability density in the system are globally balanced \cite{weigert},  whereas breaking the  $\mathcal{P}\mathcal{T}$ symmetry causes the system to exhibit accumulation or depletion of density within the system \cite{ortega2020spectral, ortega2021spectra}.  \\ 
 
Regarding the scattering aspects of the non-Hermitian $\mathcal{P}\mathcal{T}$-symmetric potential, it turns out that the transmission and reflection coefficients do not obey the usual unitary rule. Instead, other ``conservation rules'' have been established \cite{Ge, Lin, CANNATA2007397, AHMED2001343, Mostafazadeh_2014}, which we discuss in detail in Section 3. Also, it has been reported that these coefficients may present anomalous behaviour, i.e. they can take values greater than one  \cite{AHMEDHandedness}, or even be infinite at certain values in the continuous spectrum \cite{Spectralsingularities, Chaos}.  \\

In this work, we study an extension of the 
$\mathcal{P}\mathcal{T}$-symmetric square potential well presented by Znojil \cite{pozo2} and Lévai \cite{levai2018finite}. Yet, unlike those systems, ours exhibits a broken $\mathcal{P}\mathcal{T}$-symmetry phase. Our system consists of a finite $\mathcal{PT}$-symmetric complex potential well  and a delta-potential at the origin modulated by a (real) coupling constant $\lambda$. This enables the system to exhibit a complexification of its eigenvalues as a function of this parameter, thereby displaying the characteristics specific to this phenomenon, such as the appearance of exceptional points.   Qualitatively similar phenomena occur in a related discrete Hamiltonian \cite{Barashenkov}. Also, as mentioned above, a central feature of $\mathcal{PT}$-symmetric systems is the the presence of fluxes driven by the gain and loss of density and energy arising from the imaginary part of the potential, even in the $\mathcal{PT}$-symmetric stationary states. Thus, the characterization of the fluxes within these systems is crucial to understanding their behavior. In view of this, in addition to its spectral properties, we also analyze the transport properties of probability and energy within the system. Specifically, we compute and study the behavior of the density currents for bound and scattering states, as well as the associated energy densities and their respective fluxes. Also related to the transport properties of the system, we study the energy densities and energy fluxes. Since the system is solvable, we obtain closed form expressions for two energy densities definitions, and discuss their relevance in connection with the probability density and flux of probability.\\

The outline of the paper is as follows: in Sec.~\ref{sec:model} we present the  solutions of the respective Schr\"odinger equations for bound and scattering states. For the bound states we discuss the spectra as a function of the parameters of the system. For scattering states, we find the reflection and transmission coefficients and we analyze their behavior.  In Sec.~\ref{sec:transportprop}, first we discuss the continuity equations for the different quantum densities that are studied. Subsequently, the respective flows are calculated for this model for both the bound and scattering states. These fluxes are evaluated at the edges of the well to determine the efficiency of transport through the system. Finally, our results are summarized in Sec.~\ref{sec:summ}. 

%and we characterize the efficiency of transport of the quantities already mentioned. 

\section{The model and its solutions}
\label{sec:model}
We consider a finite, both in length and depth,
$\mathcal{P}\mathcal{T}-$symmetric square potential well \cite{pozo2, levai2018finite} in an infinite system, and add an extra term  proportional to a Dirac delta function. This delta-potential acts as a barrier that, as the parameter associated with it increases, there comes a point where the transport from the source to the drain is no longer efficient, leading to a breakdown of the $\mathcal{P}\mathcal{T}-$symmetry and the appearence of exceptional points. Specifically, we consider the following time independent Schrödinger equation
\begin{equation}
\left( -\frac{\hbar^2}{2m}\frac{\partial ^2}{\partial x^2 }+V(x)+\lambda \delta(x)\right) \psi(x)=E\psi(x), 
\label{eqn: sed}
\end{equation}
where $\lambda$ {is a real constant} (the strength of the delta-potential) and $V(x)$ is 
the piecewise potential, 
\begin{equation}
    V(x)=\begin{cases} 
      0, & |x|<b,  \\
      V_0+ i V_I, &  x\leq -b, \\
      V_0- i V_I, & x\geq b. 
   \end{cases}
   \label{eqn:pot}
\end{equation}

In this system, the wave function is subject to the boundary (matching) conditions
\begin{equation}
    \begin{array}{cc}
    \psi(-b^{-})=\psi(-b^{+}), &\hspace{1.65cm} \psi'(-b^{-})=\psi'(-b^{+}), \\
    \psi(b^{-})=\psi(b^{+}), &\hspace{1.7cm} \psi'(b^{-})=\psi'(b^{+}), \\
    \psi(0^{-})=\psi(0^{+}), & \psi'(0^{+})-\psi'(0^{-})=\Lambda\psi(0),
    \end{array}
    \label{eq:boundcond}
\end{equation}
where we have defined $\Lambda=\frac{2m \lambda}{\hbar^2}$ and the signs $\pm$ denote the right or left limit, respectively, of the wavefunction at a given $x$. The matching conditions at the origin allow us to identify two types of basis functions:  even (e) and odd (o) \cite{deltacoupling} 
\begin{align}
      &\psi_{e}(x)\propto\cos(kx)+\frac{\Lambda}{2k}\text{sgn}(x)\sin(kx),   \\
    &\psi_{o} (x) \propto \sin(kx), 
\end{align}
where $k=\sqrt{\frac{2m E}{\hbar^2}}$. Thus, inside the well, the $\mathcal{P}\mathcal{T}-$symmetric wave function of the system can be written as a linear combination of the even and odd basis functions as~\cite{pozo2}
 \begin{equation}
     \psi(x)=C_1\psi_{e}(x)+ iC_2\psi_{o}(x), 
 \end{equation}
 with $C_1, C_2$ real constants. 
 The complete solution of the Sch\"odinger equations is
\begin{equation}
    \psi (x)=\begin{cases} 
    C_1\left(\cos(kx)+\frac{\Lambda}{2k}\text{sgn}(x)\sin(kx)\right)+ iC_2\sin(kx),  & -b<x<b, \\
       A_{1}e^{\alpha  x}+A_{2}e^{-\alpha x}, & \hspace{0.25cm} x\leq -b, \\
       B_{1}e^{\Tilde{\alpha}x}+B_{2}e^{-\Tilde{\alpha}x}, & \hspace{0.25cm} x\geq b,
   \end{cases}
   \label{eq:psigensol}
\end{equation}
where $\alpha= \sqrt{v_0+iv_I-k^2}$, $\tilde{\alpha}= \sqrt{v_0-iv_I-k^2}$, $v_I=\frac{2m V_I}{\hbar^2}$ and $v_0=\frac{2m V_0}{\hbar^2}$. While in principle $k$ might be complex,
in the  $\mathcal{P}\mathcal{T}-$symmetric phase $k$ is real and $\Tilde{\alpha}= \alpha^*$.
 Alternatively, we can write the real $(\alpha_R)$ and imaginary $(\alpha_I)$ part of $\alpha$ as
 \begin{align}\label{alfaR}
     \alpha_{R}=\sqrt{\frac{\sqrt{\left(v_0-k^2\right)^2+v_I^2}+(v_0-k^2)}{2}}, \\
     \label{alfaI}
     \alpha_I= \sqrt{\frac{\sqrt{\left(v_0-k^2\right)^2+v_I^2}-(v_0-k^2)}{2}}.
\end{align}

\subsection{Discrete spectrum and bound states}
\begin{figure}[h]
    \centering
           \includegraphics[scale=1.5]{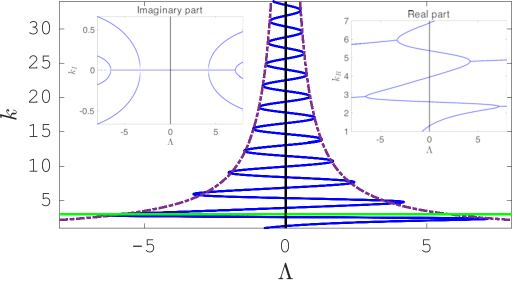}
        \caption{The blue curves represent the pseudo-momentum $k$ for the bound states as a function of $\Lambda$ with $v_0=9$, $v_I=15$ and $b=1$. These curves end at exceptional points at values bounded by $\kappa(\Lambda)$ in Eq.(\ref{eq:curve}) (purple dashed line) at large values of $k$ ($k>>\sqrt{v_0}$ solid green line).  
In the insets one can observe the complexification of the spectrum of the first eight states as a function of the parameter $\Lambda$, here $k_R$ and $k_I$ denote the real and imaginary parts of  $k$.}
 \label{fig:enter-label}
\end{figure}

The bound states are the normalizable solutions of the Schr\"odinger equation, therefore the wave function must vanish as $|x|\to \infty$. Taking  the real parts of $\alpha, \Tilde{\alpha}$ positive and making the coefficients $A_2,B_1=0$ in Eq. (\ref{eq:psigensol}), we have an asymptotically vanishing wave function 
\begin{equation}
\psi(x)=\begin{cases} 
   C_1(\cos(kx)+\frac{\Lambda}{2k}\text{sgn}(x)\sin(kx))+ iC_2\sin(kx),  & -b<x<b, \\
       A_{1}e^{\alpha x}, &  x\leq -b, \\
      B_{2}e^{-\tilde{\alpha}x}, & x\geq b.
   \end{cases}
   \label{eqn: bswf}
\end{equation}
The pseudo momentum $k$ can be obtained by solving its associated transcendental equation which is obtained via the computation of the determinant arising from the boundary conditions Eq.~(\ref{eq:boundcond}). The equation is
 \begin{equation}
 (k^2+\alpha\Tilde{\alpha})\Lambda+( 2k^2(\alpha+\tilde{\alpha})+(k^2-\alpha\Tilde{\alpha})\Lambda)\cos(2kb)+k\Big(2(\alpha\Tilde{\alpha}-k^2)+(\alpha+\tilde{\alpha})\Lambda\Big)\sin(2kb)=0.
   \label{eqn:eed}
 \end{equation}
In the $\mathcal{P}\mathcal{T}-$symmetric phase this equation can be written as
\begin{equation}
 \underbrace{\scriptstyle \big(2\Lambda(k\cos(kb)+\alpha_R\sin(kb))+4k(\alpha_R \cos(kb)-k\sin(kb))\big)\big(k\cos(kb)+\alpha_R\sin(kb)\big)}_{\text{Hermitian}}\scriptstyle+\underbrace{ \scriptstyle  2\alpha_I^2\sin(kb)\big( 2k\cos(kb)+\Lambda \sin(kb)\big)}_{\text{non-Hermitian}}=0. 
 \label{eqn:ptee}
\end{equation} 
The first term in this expression, which we denote as the "Hermitian" part, factors into two independent equations corresponding to the two types of states expected in that case. These states can be identified as $\Lambda$-dependent (even) or -independent (odd) states. In the $\mathcal{P}\mathcal{T}$-symmetric system, the imaginary part of the potential couples the two kind of states in a new equation that cannot be factorized. This coupling, along with the strength of the parameter $\Lambda$ associated with the Dirac delta, allows for the existence of exceptional points (Fig.~\ref{fig:enter-label}). As the parameter $\Lambda$ increases, the states coalesce, breaking the $\mathcal{P}\mathcal{T}$-symmetry. In this figure, for reference, we have included the solutions for $k$ when the delta-potential is absent ($\Lambda = 0$) in gray lines. In this case, there is no $\mathcal{PT}$-symmetry breaking, as noted in \cite{levai2018finite}.\\

The exceptional points appear at different values of $\Lambda$ as $k$ grows. We can approximate their behavior when $k \gg \sqrt{v_0}$ as follows (for a general reference on asymptotic expansions see~\cite{benderbook1999}). Expanding $\alpha_R$ and $\alpha_I$ from Eqs. (\ref{alfaR}) and (\ref{alfaI}) for large values of $k$, we find
\begin{align}
     &\alpha_{R}\approx \frac{v_I}{2k},\\
     &\alpha_{I}\approx k\left(1-\frac{v_0}{2k^2}\right).
\end{align}
Substituting these expressions in Equation~(\ref{eqn:ptee}) and expanding to leading order in $k$ we have
\begin{equation}
     k\Lambda+( v_I \cos(2 kb) -  v_0 \sin(2 kb))=0. 
\end{equation}
This equation has solution if we take $\Lambda\equiv\chi/k$, where $\chi$ is constant, thus
\begin{equation}
   \chi +( v_I \cos(2 kb) -  v_0 \sin(2 kb))=0,
\end{equation}
and we find 
\begin{align}
\sin(2kb)=\frac{v_0\chi\pm\sqrt{v_I^2(v_0^2+v_I^2-\chi^2)}}{v_0^2+v_I^2}.
\end{align}
From the square root, we note that real solutions only occur when 
\begin{equation}
   \chi\leq  |\sqrt{v_0^2+v_I^2}|. \label{eq:curve}
\end{equation}
Hence a curve that bounds the solutions of Eq.(\ref{eqn:eed}), including  the exceptional points, for $k\gg \sqrt{v_0}$, is $\kappa(\Lambda)=\sqrt{v_0^2+v_I^2}/\Lambda$. We show this curve in Fig. 1 (dashed purple line).  \\

\begin{figure}[h]
    \centering
       \includegraphics[width=1\linewidth]{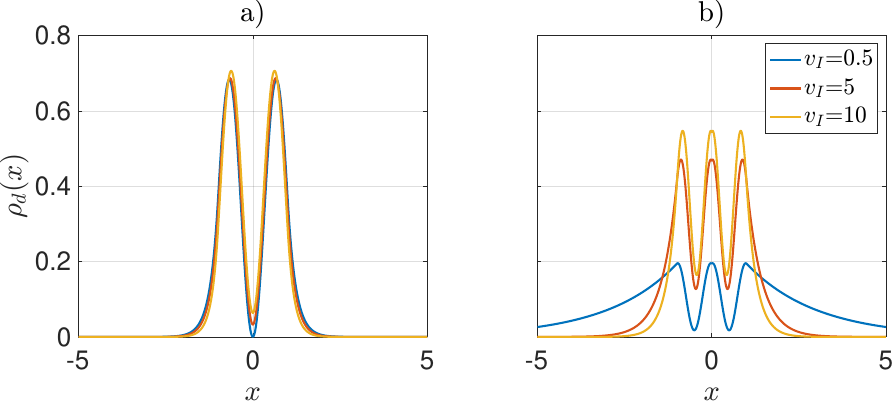}
    \caption{Density of the bound states for cases when $k^2<v_0$, a), and $k^2>v_0$, b), for some values of $v_I$  with 
 $v_0=10$,  $\Lambda =0.5$ and $b=1$. As the magnitude of the imaginary potential decreases, the bound states with $k^2>v_0$ become more extended and their magnitude decreases accordingly. }
    \label{fig:boundstates}
\end{figure}

Another important aspect that can be highlighted about this system, shown in Fig.~\ref{fig:enter-label}, is the presence of a discrete set of bound states with energy greater than $V_0$. This fact has been reported previously in similar systems \cite{Lévai_2009,levai2018finite}. Such states do not exist in the Hermitian case, which begs the question of what happens to them as the imaginary part of the potential vanishes.  To answer this question, fist we write the wave function in the $\mathcal{P}\mathcal{T}-$symmetric phase 
obtaining the coefficients explicitly 
\begin{equation}
       \psi(x)=|C_1|\begin{cases} 
 \scriptstyle  (\cos(kx)+\frac{\Lambda}{2k}\text{sgn}(x)\sin(kx)) (\alpha_R \sin (kb)+k \cos (kb))+i\alpha_I(\cos
   (kb)+\frac{\Lambda}{2k} \sin(kb)) \sin (k x ), & -b<x<b, \\
  
        e^{\alpha (x+b)} (k\cos(kb)+\alpha^*\sin(kb))\left(\cos(kb)+\frac{\Lambda}{2k} \sin(kb)\right),   & \hspace{0.3cm} x\leq -b, \\
 e^{-\alpha^*(x-b)} (k\cos(kb)+\alpha\sin(kb))\left(\cos(kb)+\frac{\Lambda}{2k} \sin(kb)\right),   & \hspace{0.3cm} x\geq b, 
   \end{cases}
   \label{eqn:deltawavefunction}
\end{equation}
where the remaining constant $C_1$ accounts for the wavefunction normalization. With this expression we compute the probability density $\rho_D(x)=|\psi(x)|^2$, and using Eq.~(\ref{eqn:eed}) we obtain 
\begin{equation}
    \rho_D(x)=|C_1|^2\begin{cases} 
\scriptstyle (\cos(kx)+\frac{\Lambda}{2k}\text{sgn}(x)\sin(kx))^2 (\alpha_R \sin (kb)+k \cos (kb))^2+\alpha_I^2( \cos
   (kb)+\frac{\Lambda}{2k} \sin(kb))^2 \sin^2 (k x ), & -b<x<b, \\
  
        e^{2\alpha _R(x+b)}  k  \Big(k \cos(kb) + \alpha_R \sin(kb)\Big)  \Big(  \cos(kb) + \frac{\Lambda}{2k} \sin(kb)\Big),  & \hspace{0.3cm} x\leq -b, \\
 e^{-2\alpha_R(x-b)} 
 k\Big(k \cos(kb) + \alpha_R \sin(kb)\Big)  \Big(  \cos(kb) + \frac{\Lambda}{2k} \sin(kb)\Big), & \hspace{0.3cm} x\geq b.
   \end{cases}
 \end{equation}
Using the normalization condition~\footnote{There are other norms used in these systems \cite{levai2018finite}, which has been established by Bagchi \cite{Bagchi} and is related to the $\mathcal{P}\mathcal{T}$-symmetric inner product proposed by Bender \cite{Bender3}. Nevertheless, this norm is not necessarily positive, which is troublesome \cite{bender2019pt}. For other systems analogous to ours, the normalization constant has been calculated using the $\mathcal{C}\mathcal{P}\mathcal{T}$ inner product \cite{BenderCproduct}, which is a positive norm,  however, this normalization constant has only been possible to calculate it exactly for discrete systems \cite{garmon2015bound} and only perturbatively for continuous systems \cite{Mostafazadeh1}.}
\begin{equation}
 \displaystyle \int_{-\infty}^{\infty} \rho_d(x)dx=1,
\end{equation}
we obtain  
\begin{equation}
 |C_1|^2=\frac{4\alpha_R k^3}{D(k)}, 
  \label{eq: nr}
\end{equation}
where $D(k)$ is given by 
\begin{equation}
\begin{split}
     D(k)=& \alpha_R\Bigg[2\sin(kb)\Big((4k^2-\Lambda^2)\cos(kb)+4k\Lambda\sin(kb)\Big)(k\cos(kb)+\alpha_R\sin(kb))^2\\&
     -4k^2\alpha_I^2\sin(2kb)\left(\cos(kb)+\frac{\Lambda}{2k}\sin(kb)\right)^2 \\& +
     2kb\Bigg((4k^2+\Lambda^2)(k\cos(kb)+\alpha_R\sin(kb))^2+4k^2\alpha_I^2\left(\cos(kb)+\frac{\Lambda}{2k}\sin(kb)\right)^2\Bigg)\Bigg]\\&+
  8k^4  \Bigg[   \Big(k \cos(kb) + \alpha_R \sin(kb)\Big)  \left( \cos(kb) + \frac{\Lambda}{2k} \sin(kb)\right)\Bigg].
\end{split}
\end{equation}
In order to understand how these states disappear while those with energies below $V_0$ remain, we focus on the behavior of the normalization constant Eq.~(\ref{eq: nr}) as $v_I\to 0$. Taking the limit of small $v_I$ in Eq. (\ref{alfaR}) we find that 
\begin{align}
     \alpha_{R}\sim \begin{cases} |v_0-k^2|^{1/2},  &  k^2\leq v_0,  \\
     \frac{v_I}{2|v_0-k^2|^{1/2}},  &  k^2>v_0.
     \end{cases}
\end{align}
% \end{align}
Thus, as the imaginary potential decreases, $\alpha_R$, and hence $|C_1|^2$, vanish as $v_I$ decreases for $k^2>v_0$, but not for $k^2<v_0$. This reflects the fact that as the potential becomes real, the bound eigenstates with energies above $V_0$ become increasingly extended and their amplitude tends to zero to maintain normalization. This behaviour can be seen in Fig.~\ref{fig:boundstates}. In this figure, one can see in a) the probability densities for bound states  with $k^2<v_0$ and for three different values of the imaginary part of the potential $v_I = 0.5,5,10$; even if the potential increases, the wavefunction remains more or less invariant. On the other hand, b) shows the case where $k^2>v_0$, i.e. bound states above the maximum of the real part of the potential, again using the same values for $v_I$ as in a); in this case, as $v_I$ decreases, the wavefunctions become flatter.
\\

\subsection{Scattering States}

\begin{figure}[h]
    \centering
     \includegraphics[width=0.75\linewidth]{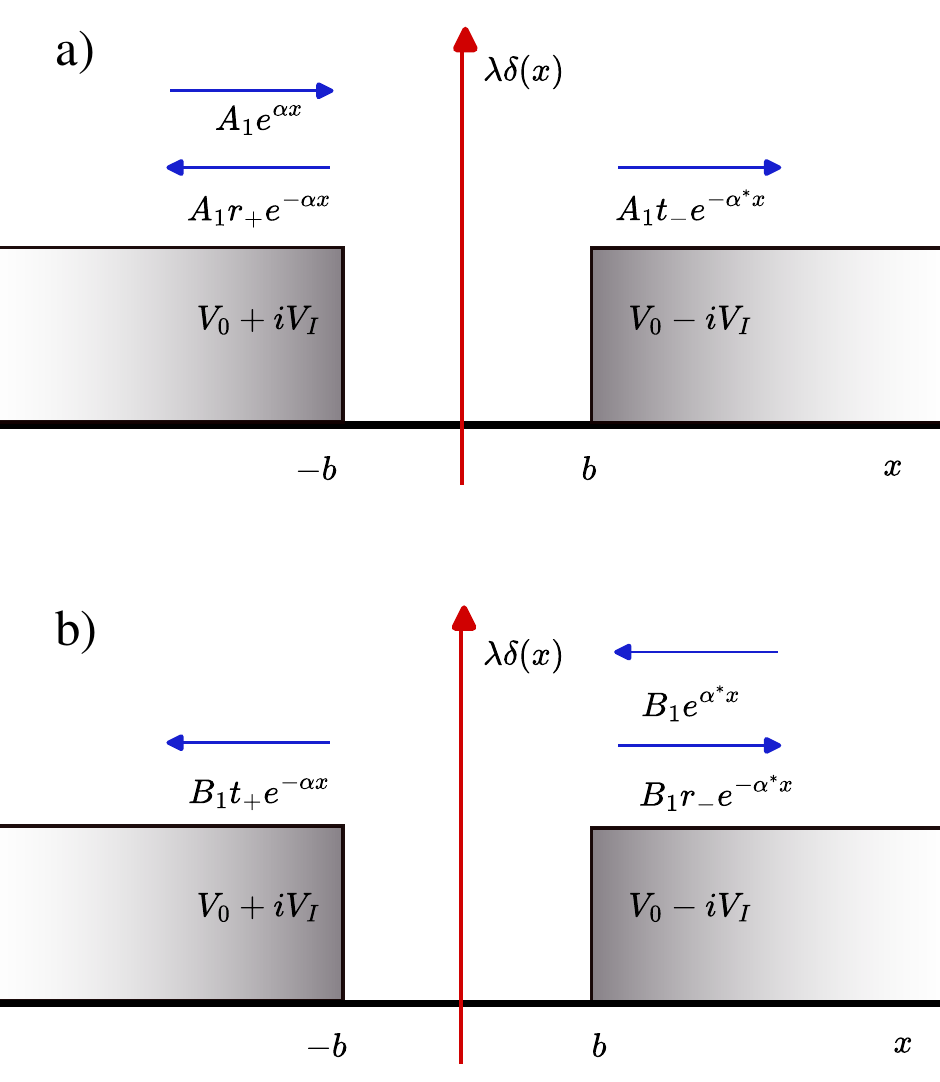}
  \caption{Scattering process. a) Left-to-right incidence and b) right-to-left incidence. The figure also illustrates the reflection $r_{\pm}$ and transmission coefficients $t_\pm$ involved in the two scattering processes.}
 \label{fig:sca}
\end{figure}

In what follows we describe the properties of scattering states that occur in this system \cite{CANNATA2007397, mostafazadeh2018scattering}.  The specific scattering processes can be seen schematically in Fig.~\ref{fig:sca}. The complete description of these setups involves both the left-to-right and right-to-left processes, as the properties of the respective reflection and transmission coefficients are intrinsically related to the scattering direction  \cite{garmon2015bound, Mostafazadeh_2014}, as we will show below.\\

First we find the solution of the Schr\"odinger equation for the left-to-right scattering case can be obtained making the coefficient $B_1=0$ in Eq.~(\ref{eq:psigensol}), thus
\begin{equation}
    \psi_{+} (x)=\begin{cases} 
    C_1(\cos(kx)+\frac{\Lambda}{2k}\text{sgn}(x)\sin(kx))+ iC_2\sin(kx),  & |x|<b, \\
       A_{1}e^{\alpha  x}+A_{2}e^{-\alpha x}, & \hspace{0.23cm} x\leq -b, \\
       B_{2}e^{-\alpha ^*x}, & \hspace{0.2cm} x\geq b. 
       \label{eqn:wf2}
   \end{cases}
\end{equation}
We define the left-to-right reflection and transmission coefficients as 
\begin{align}
    \mathbf{r}_{+}=\frac{A_2}{A_1},\\
\mathbf{t}_{-}=\frac{B_2}{A_1}. 
\end{align}
%The scattering solutions are not $\mathcal{P}\mathcal{T}-$symmetric, and due to the presence of a source and a sink, the relation $|\mathbf{r}_{+}|^2+|\mathbf{t}_{-}|^2= 1$ is not satisfied. 
%Instead we must  consider 
On the other hand, the right-to-left scattering is described by a wave function of the form (c.f. Eq.~(\ref{eq:psigensol}))
\begin{equation}
    \psi_{-} (x)=\begin{cases} 
  C_1(\cos(kx)+\frac{\Lambda}{2k}\text{sgn}(x)\sin(kx))+ iC_2\sin(kx),  & |x|<b, \\
      A_{2}e^{-\alpha x}, &  \hspace{0.25cm} x\leq -b, \\
       B_{1}e^{\alpha ^*x}+B_{2}e^{-\alpha ^*x}, & \hspace{0.25cm} x\geq b.
       \label{eqn:wf3}
   \end{cases}
\end{equation}
The right-to-left reflection and transmission coefficients are  
\begin{align}
        \mathbf{r}_{-}=\frac{B_2}{B_1},\\
\mathbf{t}_{+}=\frac{A_2}{B_1}. 
\end{align}

The complete set of reflection and transmission coefficients can be explicitly computed through the transfer matrix method (see Appendix \ref{appendix:b}), and yield
\begin{align}
 \mathbf{t}_{+} &=\scriptstyle \frac{4\alpha^*k^2}{-2k \sin (2 kb) \left( |\alpha|^2 +k^2+i\Lambda \alpha_I\right)+\cos (2 kb) \left(|\alpha|^2 \Lambda +k^2 (-4i\alpha_I+\Lambda )\right)+\Lambda  \left(k^2-|\alpha|^2\right)} e^{2ib\alpha_I}, \nonumber\\ 
 \mathbf{r}_{-} &=\scriptstyle \frac{ \left(2k \sin (2 kb) \left(\Lambda  \alpha_R - (|\alpha|^2- k^2)\right)+\cos (2 kb) \left(|\alpha|^2 \Lambda +k^2 (4\alpha_R-\Lambda )\right)-\Lambda  \left(k^2+|\alpha|^2\right)\right)}{-2k \sin (2 kb) \left( |\alpha|^2 +k^2+i\Lambda \alpha_I\right)+\cos (2 kb) \left(|\alpha|^2 \Lambda +k^2 (-4i\alpha_I+\Lambda )\right)+\Lambda  \left(k^2-|\alpha|^2\right)} e^{2b\alpha^* },\nonumber \\
 \mathbf{r}_{+} &=\scriptstyle -\frac{\left(2k \sin (2 kb) \left(\Lambda  \alpha_R+( |\alpha|^2- k^2)\right)+\cos (2 kb) \left(k^2 (4\alpha_R+\Lambda )-|\alpha|^2\Lambda \right)+\Lambda  \left(k^2+|\alpha|^2\right)\right)}{-2k \sin (2 kb) \left( |\alpha|^2 +k^2+i\Lambda \alpha_I\right)+\cos (2 kb) \left(|\alpha|^2 \Lambda +k^2 (-4i\alpha_I+\Lambda )\right)+\Lambda  \left(k^2-|\alpha|^2\right)} e^{-2b\alpha},\nonumber\\
 \mathbf{t}_{-} &= \scriptstyle-\frac{4\alpha k^2}{-2k \sin (2 kb) \left( |\alpha|^2 +k^2+i\Lambda \alpha_I\right)+\cos (2 kb) \left(|\alpha|^2 \Lambda +k^2 (-4i\alpha_I+\Lambda )\right)+\Lambda  \left(k^2-|\alpha|^2\right)}e^{2ib\alpha_I}.
 \label{eqn:coef}
\end{align}

\begin{figure}[h]
    \centering
    \includegraphics[width=0.45\linewidth]{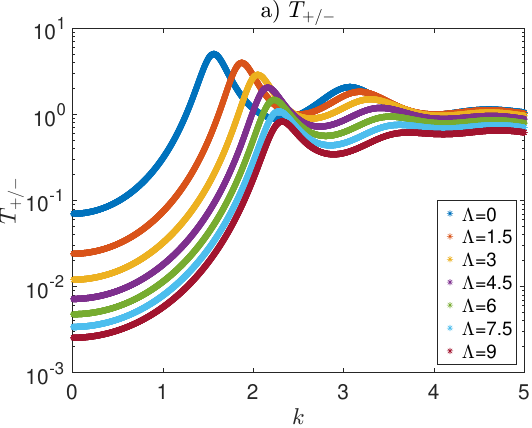}\quad
    \includegraphics[width=0.45\linewidth]{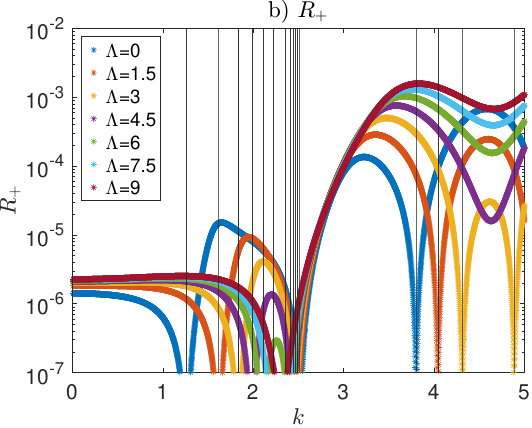}
        \vspace{1cm}
    \includegraphics[width=0.45\linewidth]{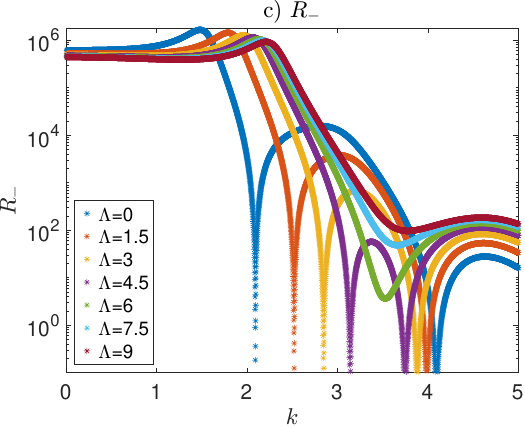}\vspace{1cm}\quad
    \includegraphics[width=0.45\linewidth]{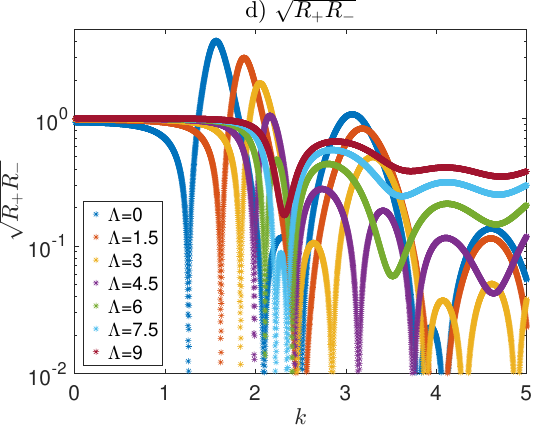}
    \caption{ a) Pseudo-transmission $T_{+/-}$, b) Left pseudo-reflection $R_+$, c) Right pseudo-reflection $R_-$ d) Module of the product of left and right pseudo-reflections as a function of $k$ for the parameter $v_I=10$, $v_I=10$ and $b=1$. The black vertical lines in b) indicate the solutions of Eq. (\ref{eqn:eed}). We can see how  the zeros of Eq. (\ref{eqn:eed}) indicate where the reflection coefficient $R_{+}(k)$ vanishes, which, due to the logarithmic scale of the figure, can be seen as the curves tending to negative infinity.}
    \label{fig:tr}
\end{figure}
From these results we can compute the pseudo-transmissions $T_\pm=|\mathbf{t_\pm}(k)|^2$ and pseudo-reflections $R_\pm=|\mathbf{r_\pm}(k)|^2$, which are shown in Fig.~\ref{fig:tr}. We note that the left and right transmission are equal,  $|\mathbf{t}_{+}(k)|^2=|\mathbf{t}_{-}(k)|^2$. Actually, the pseudo-transmissions and pseudo-reflection can show anomalous behavior, taking values greater than one as can be seen in Fig.~\ref{fig:tr}. On the other hand,  the left reflectivity ($R_{+}$) vanishes at the bound states, as expected. This can be observed in Fig.~\ref{fig:tr}  
or explicitly in the numerator of the expression for $\mathbf{r}_{+}(k)$ in Eq. (\ref{eqn:coef}). Indeed, when the left reflectivity coefficient is zero, the wave function Eq. (\ref{eqn:wf2}) takes the form of the bound state wave functions given in Eq. (\ref{eqn: bswf}). 
Fig.~\ref{fig:tr} also shows that there is a set of discrete values of $k$ at which $R_-$ vanishes, which correspond to non integrable reflection-less states, i.e. states for which the non vanishing terms in  Eq. (\ref{eqn:wf3}) diverge as $|x|\rightarrow \infty$.\\

Finally, it can also be verified that the reflection and transmission coefficients satisfy the following identities reported in general for $\mathcal{P}\mathcal{T}-$symmetric systems \cite{Ahmed_2012,Mostafazadeh_2014}
\begin{align}
    \mathbf{r}_{\pm}(-k)= \mathbf{r}_{\pm}(k), \\
    \mathbf{t}_{\pm}(-k)= \mathbf{t}_{\pm}(k).
\end{align}
Indeed, the system satisfies the "generalized unitary relation" \cite{Ge, Mostafazadeh_2014} 
\begin{equation}
|\mathbf{t}_{\pm}(k)|^2  \pm  |\mathbf{r}_{-}(k)\mathbf{r}_{+}(k)|=1.
\label{eqn:unitary}
\end{equation}
The sign is chosen according to whether the quantity, $1-T= 1-|\mathbf{t}_{+}(k)|^2 $ \cite{garmon2015bound} is positive or negative: in the regions where the pseudo-transmission presents an anomalous behavior $T> 1$, the negative sign will be chosen, otherwise, where $T<1$ one chooses the positive one.

\section{Transport properties }
\label{sec:transportprop}
In this section we focus on the analysis of the transport properties of the system. Specifically, we center our attention on the probability and energy densities and their respective fluxes. However, to do so, it is convenient to discuss briefly two different definitions for the energy density, as these yield different continuity equations, and only then analyze their transport properties in the context of our specific problem.

\subsection{Probability density, energy density and their continuity equations}
\label{sec:probden_gen}
In general, we consider a particle in an arbitrary complex potential
\begin{equation}
V(x)=V_{R}(x)+iV_{I}(x). 
\label{eqn:complex_potential}
\end{equation}
The particle is described by the wavefunction $\psi(x,t)$ for which the probability density is $\rho_d(x,t)=|\psi(x,t)|^2$. This probability density satisfies the
(standard) continuity equation, which can be derived by computing the time derivative of the probability density $\rho$ and 
the Sch{\"o}dinger equation, along with its complex conjugate, and taking into account the complex potential in Eq.~(\ref{eqn:complex_potential}). Several examples of such equation include for instance \cite{Molinas, KOCINAC2008191, Ernzerhof, Elenewski, Ahmed2001, Zhang2021}. The resulting continuity equation is
\begin{equation}
\frac{\partial \rho_d(x,t)}{\partial t}+ \frac{\partial }{\partial x} J_d(x,t)=Q_d(x,t),
\label{eqn:ced}
\end{equation}
where the probability current is
\begin{equation}
    J_d(x,t)=\frac{\hbar}{2m i}\left[\psi^*(x,t) \frac{\partial}{\partial x} \psi(x,t)-\psi(x,t)\frac{\partial}{\partial x} \psi^*(x,t)\right],
    \label{eqn:cenm}
\end{equation} and 
\begin{equation}
    Q_d(x,t)=2\frac{V_{I}(x)\rho_d(x,t) }{\hbar},
\end{equation} 
accounts for the source/drain terms. Thus, under this perspective, the imaginary part of the potential is interpreted as giving rise to a source or sink of probability~\cite{Molinas}. It is important to mention that a self-consistent conservation law has been proposed for the norm defined by the $\mathcal{P}\mathcal{T}$ inner product \cite{Bagchi,Japaridze_2002}. However, this norm usually presents certain drawbacks, so for the purposes of this work, we will not adopt it. Instead, we will approach the transport properties of the probability density from the perspective of the continuity equation (\ref{eqn:ced}).\\

Another quantity that satisfies a continuity equation in quantum mechanics, and is less explored, is the energy density \cite{Mathews, LudovicoPRB2014, Arvizu_2023}. Unlike the definition for the probability density, the definition for the energy density in quantum mechanics is not unique, different definitions yield different results and therefore may give different values for the flux of energy in a quantum system. An initial attempt to define the energy density could be simply as
\begin{equation}
    \rho^E(x,t)=\psi^*(x,t)H \psi(x,t).
\end{equation}
However, this expression is not necessarily real, which makes its interpretation as a physical density troublesome. However, for $\mathcal{PT}-$symmetric states
\begin{eqnarray}
   \int  V_{I}(x)\rho_d(x,t)dx=0.
   \label{eqn:a}
 \end{eqnarray}
so $V_I$ does not contribute to the expectation value of the energy. Instead, we note that the integrand of the previous equation is proportional to $Q_d(x,t)$ \cite{Molinas,AHMED2001343, Wahlstrand}, therefore, equation~(\ref{eqn:a}) can be interpreted as saying that in $\mathcal{PT}-$symmetric systems gain and loss of density is globally balanced \cite{weigert,ortega2020spectral}. Thus, $\mathcal{PT}-$symmetric stationary states must have sustained currents carrying the probability density and energy density that are being input to where they are being removed.\\
  
Equation~(\ref{eqn:a}) allows us to obtain a first acceptable definition of the energy density ($\rho_1^E$) for the $\mathcal{PT}-$symmetric phase as \begin{equation} \rho^E_1(x,t)= \frac{1}{2}\big(\psi^* (x,t)H \psi(x,t) + \psi(x,t) H^\dagger \psi^*(x,t)\big). \end{equation} 
This definition yields a real energy density, but has the drawback of not being necessarily positive even for non-negative potentials.
Alternatively, another definition of energy density is
\begin{equation} \rho^E_2 (x,t)= \frac{\hbar^2}{2m}\left|\frac{\partial}{\partial x} \psi(x,t)\right|^2 + V_{R}(x)|\psi(x,t)|^2. \end{equation} 
This expression also integrates to the total expected energy of the system, as can be checked by integrating by parts and using that the wave function vanishes at infinity, and has a positive kinetic energy density. The physical relevance, of either expression is an open question \cite{stepanyanQuantum2024, Mathews,  Arvizu_2023}, so in what follows, we study the properties for both expressions.\\

Following the procedure shown in \cite{Arvizu_2023}, it is possible to derive continuity equations for the two definitions of energy density,
\begin{align}
\frac{\partial \rho^E_1(x,t)}{\partial t}+\frac{\partial}{\partial x} J_{1}^E(x,t)= Q^E(x,t)
, \\
\frac{\partial \rho^E_2 (x,t)}{\partial t} +\frac{\partial}{\partial x} J_{2}^E(x,t) =Q ^E(x,t),
\end{align}
where $Q^E(x,t)
=2\frac{V_{I}(x,t)\rho^E_2(x,t)}{\hbar}$ is the same in both cases, and the respective energy density fluxes $J_1^E$ and  $J_2^E$ are given by
\begin{align}
J_1^E(x,t)&=\frac{\hbar}{4m i} \Big[
\psi^*(x,t)\frac{\partial}{\partial x}[H\psi(x,t)]-\psi(x,t)\frac{\partial}{\partial x}[H\psi(x,t)]^*
\label{eq:flux1}
\\\nonumber&-\Big(\frac{\partial}{\partial x}\psi^*(x,t)\Big)H\psi(x,t)+\Big(\frac{\partial}{\partial x}\psi(x,t)\Big)[H\psi(x,t)]^*\Big],\\
J_2^E(x,t)&=\frac{\hbar}{2 m i}\left[\left(\frac{\partial}{\partial x}\psi(x,t)\right)[H\psi(x,t)]^* - \left(\frac{\partial}{\partial x}\psi (x,t)^*\right)H\psi(x,t) \right],
\label{eq:flux12}
\end{align}
which are similar to the energy density fluxes that appear in the Hermitian case \cite{Mathews, LudovicoPRB2014, Arvizu_2023}. 

In Sec.~\ref{sec:bst} we first analyze the probability and energy transport properties of the bound states. Subsequently, in Sec.~\ref{sec:sctst}, we analyze the transport properties of the scattering states.
\subsubsection{Bound states}
\label{sec:bst}

\begin{center}
    \begin{figure}
    \includegraphics[width=0.45\linewidth]{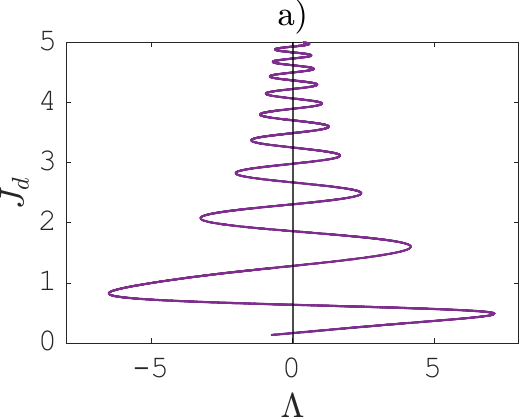}
    \hspace{0.5 cm}
    \includegraphics[width=0.45\linewidth]{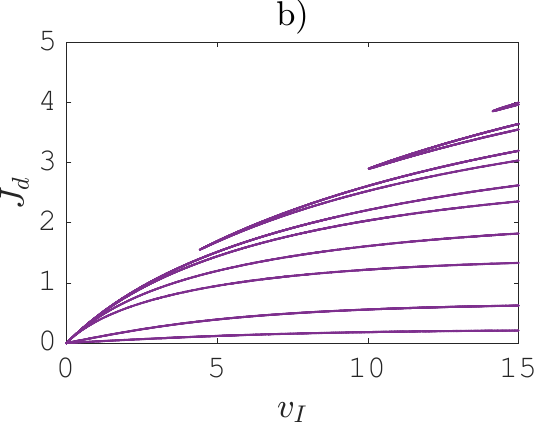}
    \caption{Plot of the flux inside the well for each of the bound states corresponding to each value of $\Lambda$ (with $v_0=9$ and $b=1$) . In a) in which we keep $v_I=15$ constant, we observe that the flux is alternating between increasing and decreasing with $\Lambda$, up to the exceptional point at which the states coalesce. On the other hand, in b) where we keep $\Lambda=0.5$ fixed, we see that $J_d(0)$ is always a monotonously increasing function of $v_I$.}
    \label{fig:flux}
\end{figure}
\end{center}

First we compute the probability density flux in the $\mathcal{P}\mathcal{T}-$symmetric phase, i.e.
 \begin{equation}
     J_d(x)=|C_1|^2\frac{\alpha_I \hbar}{m}k  \Big(k \cos(kb) + \alpha_R \sin(kb)\Big)  \Big(  \cos(kb) + \frac{\Lambda}{2k} \sin(kb)\Big)\times \begin{cases} 
 1, & -b<x<b, \\
        e^{2\alpha _R(x+b)},   & \hspace{0.25cm} x\leq -b, \\
 e^{-2\alpha_R(x-b)}, & \hspace{0.25cm} x\geq b. 
   \end{cases}
 \end{equation}
where $|C_1|^2$ is given in Eq.(\ref{eq: nr}). We can see in this expression that the flux on the side of the source ($x\leq b$) increases as it approaches the edge of the well. Once there, it remains constant across the well, and it decreases on the other side of the well.  Because of this, the flux inside the well can be viewed as the flux across the loss-gain interface, which depends on the parameters $v_I$ and $\Lambda$. These two parameters, together with $v_0$ and $b$, through Eq. (\ref{eqn:eed}), determine the values of $k$ corresponding to the spectrum of bound states and, consequently, the flux corresponding to each of these states. In the left panel of  Fig. \ref{fig:flux}, the flux behaves in a manner analogous to $k$ (Fig. \ref{fig:enter-label}), when  $v_I$, $v_0$ and $b$ are kept constant, while $\Lambda$ is varied. In this case, both $k$ and $J_d(0)$ corresponding to successive states either increase or decrease  monotonously with $\Lambda$. These two values of $k$ and $J_d(0)$ eventually coalesce at the exceptional point ~\cite{BarashenkovNJP2016}.  In the right panel, the flux's dependence on the parameter $v_I$ is shown, with $\Lambda$ and $v_0$ held constant. We observe that the flux is an increasing monotonic function of $v_I$.  This is consistent with systems similar  to ours, such as the  $\mathcal{PT-}$symmetric square well without a delta function potential, that never shows a broken $\mathcal{PT-}$symmetry phase as a function $v_I$~\cite{levai2018finite}, and there is no saturation of probability. In our system the symmetry breaking is due to the combination of $v_I$ and $\Lambda$, so the role attributable to the source in the systems studied in \cite{BarashenkovNJP2016} can be attributed in principle to the role played by the delta potential as a barrier.

 As a consequence of the stationary Schr\"odinger equation, the energy density fluxes in Eqs.~(\ref{eq:flux1}) and (\ref{eq:flux12}) are
\begin{eqnarray}
    J_1^E(x)= J_2^E(x)=\frac{\hbar^2 k^2}{2m} J_d(x).
\end{eqnarray}
The source/drain terms of probability density and energy density are given by
\begin{eqnarray}
     Q_d(x)=\frac{2V_I |C_1|^2}{\hbar}\begin{cases} 
 0, & |x|<b, \\
   \scriptstyle e^{2\alpha _R(x+b)}  k\Big(k \cos(kb) + \alpha_R \sin(kb)\Big)  \Big(  \cos(kb) + \frac{\Lambda}{2k} \sin(kb)\Big), & \hspace{0.25cm} x\leq -b, \\
\scriptstyle - e^{-2\alpha_R(x-b)} k \Big(k \cos(kb) + \alpha_R \sin(kb)\Big)  \Big( \cos(kb) + \frac{\Lambda}{2k} \sin(kb)\Big), & \hspace{0.26cm} x\geq b,
   \end{cases}
\end{eqnarray}
and
\begin{equation}
    Q_E(x)=\frac{\hbar^2k^2}{2m } Q_d(x).
\end{equation}
The integral of this quantity over the entire system vanishes, which means there is no net gain nor loss of density or energy, as expected in  the \(\mathcal{P}\mathcal{T}\)-symmetric phase. \\
 
On the other hand, the energy density $\rho_2^E$ is 
\begin{equation}
\rho_2^E(x)=
    |C_1|^2\frac{\hbar^2}{2m} \begin{cases} 
\scriptstyle k^2
\Big[\big(\Lambda\delta(x)+(-\sin(kx)+\frac{\Lambda}{2k}\text{sgn}(x)\cos(kx))^2 \big)(\alpha_R \sin (kb)+k \cos (kb))^2&\\
\scriptstyle+\alpha_I^2(\cos
   (kb)+\frac{\Lambda}{2k} \sin(kb))^2 \cos^2 (k x )
\Big], 
& -b<x<b, \\
\scriptstyle  e^{2\alpha _R(x+b)}   \left(2\alpha_R^2+k^2\right)k \Big(k \cos(kb) + \alpha_R \sin(kb)\Big)  \Big(  \cos(kb) + \frac{\Lambda}{2k} \sin(kb)\Big),  & \hspace{0.27cm} x\leq -b, \\
\scriptstyle e^{-2\alpha_R(x-b)} 
 \left(2\alpha_R^2+k^2\right)k \Big(k \cos(kb) + \alpha_R \sin(kb)\Big)  \Big(  \cos(kb) + \frac{\Lambda}{2k} \sin(kb)\Big), & \hspace{0.3cm} x\geq b. 
   \end{cases}
 \end{equation} 
 It is interesting to see that for the bound states, the energy density flux is the same for both definitions, while the energy densities are markedly different. The behavior of $\rho_1^E$ is essentially the same as that of the probability density, while the potential induces a discontinuity in the energy density $\rho_2^E$. 
\subsubsection{Scattering states}
\label{sec:sctst}
In the scattering case, starting with the left to right incidence we have that the flux  in the source side can be written as 
\begin{equation}
    J_{d_{+}}(x)=%\underbracket[0.8pt]{
    \frac{|A_{1}|^2 \hbar}{m}\left[ \alpha_I \left( e^{2\alpha_R x}
    %}_{J_{D_{\mathbf{i}}}}\underbracket[0.8pt]{
    -|\mathbf{r}_{+}(k)|^2 e^{-2\alpha_Rx}\right)
    %}_{J_{D_{\mathbf{r}}}}+\underbracket[0.8pt]{
+i\alpha_R %\frac{\hbar}{m}(A_{1}^*b^{+}
\big(\mathbf{r}_{+}(k)e^{-2i\alpha_I x}-\mathbf{r}^*_{+}(k)e^{2i\alpha_Ix}\big)
%A_{1}b^{+}^* 
\right]%}_{J_{D_{\mathbf{i},\mathbf{r}}}}
\label{eqn: lflux}
\end{equation}
and the flux in the right side is 
\begin{equation}
    J_{d_{-}}(x)=
 \frac{| A_{1}|^2   |\mathbf{t}_{-}(k)|^2 \hbar \alpha_I e^{-2\alpha_R x}}{m}. 
 \label{eqn: rflux}
\end{equation}
Similarly, in the case of right to left incidence, the density flux on the right side of the well is
\begin{equation}
    J_{d_{-}}(x)= \frac{ |B_{1}|^2  \hbar}{m}\left[%\underbracket[0.8pt]{
   \alpha_I \left(|\mathbf{r}_{-}(k)|^2 e^{-2\alpha_Rx}
    -e^{2\alpha_R x}\right)
  +i\alpha_R(\mathbf{r}_{-}(k)e^{ 2i\alpha_I x}-\mathbf{r}^*_{-} (k)e^{-2i\alpha_I) x})\right],%
\end{equation}
and the flux on the left side, i.e. the flux for the transmitted wave is 
\begin{equation}
    J_{d_{+}}(x)
    %J_{D_{\mathbf{t}}}(x)
    =
   \frac{| B_1|^2 |\mathbf{t}_{+}(k)|^2 \hbar\alpha_I e^{-2\alpha_R x}}{m}.
\end{equation}
  Evaluating the fluxes at the edges of the well, we have the following conservation law 
  \begin{equation}
      J_{d_{+}}(-b)=J_{d_{-}}(b).
      \label{eqn:cond}
  \end{equation}
As in the case for bound states, we have efficient transport, in the sense that the total gain and loss in the system balance exactly, also in this case. Relation~(\ref{eqn:cond}) is also interesting because from it one can derive the generalized unitary condition Eq.(\ref{eqn:unitary}). In order to do so, we start with condition~(\ref{eqn:cond}) and we evaluate Eq.~(\ref{eqn: lflux}) in $x=-b$, this yields
\begin{equation}
\begin{split}
J_{d_{+}}(-b)=\scriptstyle \alpha_I\frac{|A_{1}|^2 \hbar}{m}  e^{-2\alpha_R b}
\Bigg[1 -&\scriptstyle\left(2k \sin (2 kb) \left(\Lambda  \alpha_R - (|\alpha|^2- k^2)\right)+\cos (2 kb) \left(|\alpha|^2 \Lambda +k^2 (4\alpha_R-\Lambda )\right)-\Lambda  \left(k^2+|\alpha|^2\right)\right)\times \\&\scriptstyle\frac{\left(2k \sin (2 kb) \left(\Lambda  \alpha_R+( |\alpha|^2- k^2)\right)+\cos (2 kb) \left(k^2 (4\alpha_R+\Lambda )-|\alpha|^2\Lambda \right)+\Lambda  \left(k^2+|\alpha|^2\right)\right)}{((k^2 - |\alpha|^2) \Lambda - (k^2 + \
|\alpha|^2) (-\Lambda \cos(2 b k) + 
    2 k \sin(2 b k)))^2+(2 k \alpha_I (2 k \cos(2 b k) + \Lambda\sin(2 b k)))^2} \Bigg].
 \end{split}
\end{equation}
From Eq.~(\ref{eqn:coef}) we can write
\begin{equation}
    J_{d_{+}}(-b)=\alpha_I
    \frac{|A_{1}|^2 \hbar}{m}  e^{-2\alpha_R b}\Bigg[1 -|\mathbf{r}_{+}(k)\mathbf{r}_{-} (k)|\Bigg].
\end{equation}
Finally, equating the last equation with Eq.~(\ref{eqn: rflux}) we obtain again the unitary relation Eq.~(\ref{eqn:unitary}), i.e.
\begin{equation}
|\mathbf{t}_{\pm}(k)|^2  \pm  |\mathbf{r}_{-}(k)\mathbf{r}_{+}(k)|=1.
%%\label{eqn:unitary}
\end{equation}
In the case of Hermitian systems, this relation reduces to the standard probability conservation $|\mathbf{t}_{\pm}(k)|^2 + |\mathbf{r}_\pm(k)|^2 = 1$.
%in our case probability is not conserved in general.

\section{Summary and conclusions}
\label{sec:summ}
In this article, we investigate the spectral and transport properties of both energy and density in a \(\mathcal{P}\mathcal{T}\)-symmetric system. For this purpose, we analyzed a relatively simple system which possesses both bound and scattering states. Specifically, the system we considered is a \(\mathcal{P}\mathcal{T}\)-symmetric finite potential well with a delta-potential at the origin of strength $\lambda$. It is worth noting that without the delta potential, the eigenvalues of the system remain real \cite{levai2018finite} for any value of $V_I$, the imaginary part of the potential.
However, the combination of \(\lambda\) and \(V_I\) allows the complexification of the eigenvalues and the breaking of the \(\mathcal{P}\mathcal{T}\) symmetry. This can be explained through the equation for the pseudo momentum $k$, there the imaginary part of the potential couples the values corresponding to eigenfunctions that would be odd (and hence independent of the delta potential) and those that would be even in the abscence of $V_I$. This coupling becomes stronger as we increase the value of the parameters, finally producing exceptional points. We were also able to describe the approximate value at which the exceptional points appear in the limit \(k \gg v_0\). \\

Another interesting aspect of the spectrum of these \(\mathcal{P}\mathcal{T}\)-symmetric finite potential well systems is the presence of bound states with energies \(E > V_0\). We also saw how these states disappear when the imaginary part of the potential goes to zero, which has to occur as we approach the Hermitian case. We found that the bound states with energy \(E > V_0\) become infinitely extended causing their amplitude to vanish when the imaginary potential tends to zero, whereas those with energies \(V_0 > E\) remain localized.\\

Regarding the transport properties of the system, the transmission and reflection coefficients were calculated. As expected, the zeros of the left reflection coefficient correspond to the eigenvalues of the bound states, because the incident wave and the transmitted wave both vanish asymptotically, so at the zeros of the transmission, the scattered and bound wave functions are the same. In the reverse process, the scattering from right to left, the states that correspond to the zeros of the reflection coefficient are a discrete set of reflectionless states that diverge to infinity.\\

The transport efficiency of the system can be quantified by evaluating the density fluxes at the edges of the well, that is, in the positions of the source and drain. In these regions, we can observe that, everything that comes from the source reaches the sink, the gain and loss balance, such that there is no accumulation or depletion of probability or energy in the system. This characteristic has also been reported for tight-binding Hamiltonians with gain and loss sites \cite{ortega2020spectral,ortega2021spectra}. However, as the strength of the delta increases it acts as a kind of barrier, which modulates the flow of probability and energy, interfering with the transport efficiency of the system, and eventually leading to a break in the $\mathcal{PT}$-symmetry. In this sense, unlike other systems, the parameter that modulates the source ($V_I$) does not cause a drop in the flow that precedes the appearance of the exceptional point, as has been reported in other systems \cite{BarashenkovNJP2016}.\\

Regarding the analysis of the energy fluxes, we mention that this is a topic addressed at least since the 60s~\cite{epsteinINC1965}, and still under discussion in different instances (see~\cite{Mathews, wuJPA2009, Arvizu_2023, stepanyanQuantum2024}).
Up to now, there is no indication that the matter has been settled, in the sense that a standard definition for the energy density is still missing. We address this quantity here because we believe that experiments based on $\mathcal{PT}$-symmetric systems can be carried out in order to determine the nature of this quantity, and decide how an energy density should be defined in Quantum Physics.\\

Finally, since the system studied here can be solved exactly, it would be of interest
to use our findings in extending the analysis to some other effects which might also be of potential interest. These comprise, but are not limited	to, the case of analyzing time-dependent wavepackets (e.g. Gaussian wavepackets) and their evolution throughout the $\mathcal{PT}$ scattering system, the so-called topological energy transfer with exceptional points~\cite{XuNaturelett2016}, the study of the jamming anomaly~\cite{BarashenkovNJP2016}, etc.

\section*{Acknowledgements}

Francisco Ricardo Torres Arvizu acknowledges support from CONAHCYT scholarship number 834573.  Hernán Larralde and Francisco Ricardo Torres Arvizu acknowledges the National Autonomous University of México (UNAM) through the SUPPORT PROGRAM FOR RESEARCH AND TECHNOLOGICAL INNOVATION PROJECTS (PAPIIT) key IN103724. A. O. acknowledges support from the program ``Apoyos para la Incorporación de
Investigadoras e Investigadores Vinculada a la Consolidaci\'on Institucional de Grupos de Investigaci\'on 2023'' from CONAHCYT, México.
  \appendix
\section{Transfer matrix}
\label{appendix:b}
The matching conditions can be summarize using the transfer matrices $\mathcal{M}^{+}(k) $, $\mathcal{M}^{-}(k)$ which are defined by the relations 
\begin{equation}
        \mathcal{M}^{+}(k) \begin{bmatrix}
        A_{1}\\ A_{2}
    \end{bmatrix}=  \begin{bmatrix}
        B_{2}\\ B_{1}
    \end{bmatrix}, 
    \qquad\mathcal{M}^{-}(k) \begin{bmatrix}
        B_{2}\\ B_{1}
    \end{bmatrix}=  \begin{bmatrix}
        A_{1}\\ A_{2}
    \end{bmatrix}, 
\end{equation}
it can easily see that $\mathcal{M}^{+}(k)^{-1}=\mathcal{M}^{-}(k)$.
Thus the components of transfer matrix are 
\begin{align}
    \mathcal{M}^{+}_{1,1}= \scriptstyle \frac{e^{-2ib\alpha_I} \left( 2 k \sin (2 kb) \left(2(|\alpha| + k^2)-i\Lambda \alpha_I \right)-\cos (2 kb) \left(|\alpha|^2   \Lambda +k^2 (4 i\alpha_I+\Lambda )\right)-\Lambda(k^2-|\alpha|^2)  \right)}{4 \alpha^* k^2},\nonumber\\
\mathcal{M}^{+}_{1,2}= \scriptstyle \frac{e^{2b \alpha_R} \left(2k \sin (2 kb) \left(\Lambda  \alpha_R- (|\alpha|^2- k^2)\right)+\cos (2 kb) \left(|\alpha|^2  \Lambda +k^2 (4 \alpha_R-\Lambda )\right)-\Lambda  \left(k^2+|\alpha|^2\right)\right)}{4 \alpha^* k^2},\nonumber\\
    \mathcal{M}^{+}_{2,1}=\scriptstyle
 \frac{e^{-2b \alpha_R} \left(-2k \sin (2 kb) \left(\Lambda  \alpha_R+(| \alpha  |^2- k^2)\right)+\cos (2 kb) \left(k^2 (4 \alpha_R+\Lambda )-|\alpha|^2 \Lambda \right)+\Lambda  \left(k^2+|\alpha|^2\right)\right)}{4 \alpha ^* k^2},\nonumber\\
 \mathcal{M}^{+}_{2,2}=\scriptstyle \frac{e^{2ib\alpha_I} \left(-2k \sin (2 kb) \left(( |\alpha|^2+k^2)+i\Lambda \alpha_I \right)+\cos (2 kb) \left(|\alpha|^2  \Lambda +k^2 (-4 i\alpha_I+\Lambda )\right)+\Lambda  \left(k^2-|\alpha  |^2\right)\right)}{4 \alpha^* k^2}, 
\end{align}
Is well known that the transfer matrix (and its inverse) can be written in terms of the transmission and reflection coefficients (denoted $\mathbf{r},\mathbf{t}$ respectively), and considering the above we have \cite{mostafazadeh2018scattering}
\begin{align}
    \mathcal{M}^{+}(k)=\begin{bmatrix}
    \mathbf{t}_{-}(k)-\mathbf{r}_{+}(k) \mathbf{r}_{-}(k)/\mathbf{t}_{+}(k) & \mathbf{r}_{-}(k)/\mathbf{t}_{+}(k)\\
    -\mathbf{r}_{+}(k)/\mathbf{t}_{+}(k) & 1/\mathbf{t}_{+}(k)
\end{bmatrix},  \\
\mathcal{M}^{-}(k)=\begin{bmatrix}
1/\mathbf{t}_{-}(k)&  -\mathbf{r}_{-}(k)/\mathbf{t}_{-}(k) \\
 \mathbf{r}_{+}(k)/\mathbf{t}_{-}(k)&   \mathbf{t}_{+}(k)-\mathbf{r}_{+}(k) \mathbf{r}_{-}(k)/\mathbf{t}_{-}(k)
\end{bmatrix} .
\end{align}
Also, the determinant of this matrix in this case is
\begin{equation}
    \text{det}(\mathcal{M}^{+})=-\frac{\alpha}{\alpha^*}= \frac{ \mathbf{t}_{-}(k)}{ \mathbf{t}_{+}(k)};
\end{equation}
which means that there is no reciprocity in transmission from one or the other direction. Hence, we write the wave function of the scattering process as 
  \begin{equation}
    \psi_{+}(x)=A_{1}\begin{cases} 
   \frac{\Big( k\cos(k(b-x))+\alpha^* \sin(k(b-x)))\Big)\mathbf{t}_{-}(k)e^{-\alpha^*  b}}{k},& |x|<b, \\ 
       e^{\alpha  x}+\mathbf{r}_{+}(k)e^{-\alpha x}, &  x\leq -b,\\ 
     \mathbf{t}_{-}(k)e^{-\alpha^*x}, 
       & x\geq b,
   \end{cases}
\end{equation}
\begin{equation}
    \psi_{^{-}}(x)=B_{1}\begin{cases} 
   \frac{\Big( k\cos(k(x+b)-\alpha \sin(k(x+b)))\Big)\mathbf{t}_{+}(k)e^{\alpha  b}}{k}, & |x|<b, \\
       \mathbf{t}_{+}(k)e^{-\alpha x}, &  x\leq -b, \\
      e^{\alpha^*x}+ \mathbf{r}_{-}(k)e^{-\alpha^*  x},
       & x\geq b. 
   \end{cases}
\end{equation}

\bibliography{references.bib} 
\end{document}